
 \magnification=\magstep1

 \rightline{NO-ADRB/96/1} \smallskip
\def\gsim{\mathrel{\raise.3ex\hbox{$>$\kern-.75em\lower1ex\hbox{$\sim$}}}}
\def\lsim{\mathrel{\raise.3ex\hbox{$<$\kern-.75em\lower1ex\hbox{$\sim$}}}}

\centerline{\bf 
MASSIVE NEUTRINOS AND THE PROBLEM OF} 

  \centerline{\bf  THE DARK MATTER IN SPIRAL GALAXIES}  \smallskip 

\centerline{\bf Srdjan Samurovi\' c$^1$ and Vladan \v Celebonovi\' c$^2$} 
\medskip \centerline{\it $^1$ People's Observatory, Gornji grad 16, } 

\centerline{\it Kalemegdan, 11000 
Belgrade, Yugoslavia}
\centerline{\it e-mail: srdjanss@afrodita.rcub.bg.ac.yu}
\smallskip
\centerline{\it $^2$ Institute of Physics, Pregrevica 118, }

\centerline{\it  11080 Zemun--Belgrade, Yugoslavia} 

\centerline{\it e-mail: celebonovic@exp.phy.bg.ac.yu}

\bigskip

\leftline{\bf 1. Introduction}

\medskip
One of the most important characteristics of spiral galaxies, rotation curves 
(hereafter RCs), do not show Keplerian fall-off thus indicating  the 
presence of invisible mass component (Ashman, 1992). The question of the 
nature of the dark matter (DM) which dominates the outer parts of spirals, but 
can also be found in the regions where the ordinary matter is present 
(Persic, Salucci, Stel, 1995, hereafter PSS), may lead towards the answer to 
the important questions concerning galaxy formation, cosmology and particle 
physics. In this paper we made the attempt to show whether massive neutrinos 
could play the important role in the above mentioned ``conspiracy'' concerning 
RCs (Peebles, 1993).
\bigskip
\leftline{\bf 2. Neutrino Mass}
\medskip

The question whether neutrinos have masses is still unresolved and we present 
here the laboratory limits on the neutrino mass that follow purely from 
kinematics:      
$m_{\nu _e}\ \lsim 5\, {\rm eV},$
$m_{\nu _\mu}\ \lsim 250\, {\rm KeV},$ 
and
$m_{\nu _\tau}\ \lsim 23\, {\rm MeV}$
(Montanet {\it et al.\/}, 1994). 
Recent measurements that have been done using the LSND (Liquid Scintillator 
Neutrino Detector) (Athanassopoulos {\it et al.\/}, 1995, Hill, 1995) strongly 
suggest that neutrinos do have masses thus giving us a possibility to determine 
their place in the universe. The contribution of neutrinos to the cosmological 
density today (Primack, 1996) is:

$$\Omega _\nu ={\sum _i m_{\nu _i}\over 94h^2 {\rm eV}},$$

\noindent where the density parameter 
$\Omega _\nu ={\rho _\nu \over \rho _{\rm crit}}$, $\rho _{\rm 
crit}={3H_0^2\over8\pi G}\cong 1.88\times 10^{-29}h^2\, {\rm g\over cm^3}$. 
 $h$ is dimensionless 
parameter used in the parametrization of the Hubble constant $H_0$, 
$H_0=100\ h\, {\rm km/s\over Mpc}$ and $0.4<h<1$.
According to the well-known cosmological upper limit 
(e.g., Sarkar, 1996) we have the following constraint:

$$\sum \limits_{i=1} m_{\nu _i}  {g_{\nu _i}\overwithdelims () 2} \le 
94 {\rm eV}, \eqno (1)$$

\noindent where the sum goes over all species that were relativistic at 
decoupling i.e. $m_{\nu _i}\ \lsim 1$ MeV.

\bigskip

\leftline{\bf 2. Neutrinos in Galactic Halos}

\medskip

Perhaps the most noticeable evidence of the existence of the DM came from the 
measurements of luminosity profiles and RCs of spiral galaxies (Rubin {\it et 
al.\/}, 1985). The case of the spiral galaxy NGC3198 (``everyone's favourite'') 
is very importanrt and could lead us to some important results concerning the 
problem of the nature of the DM. Van Albada {\it et al.\/}, (1985) found that 
the amount of the DM inside the last point of the rotational curve is at least 
four times larger than the amount of visible matter. Mass-to-light ratio 
adjusted to fit the inner part of the rotational curve of this galaxy is:

$${\cal M\over  L}=5.3\ h \ {\cal M_\odot \over  L_\odot}$$

\noindent (Peebles, 1993), where ${\cal M}$ stands for the mass, ${\cal L}$ 
for the luminosity, symbol $\odot$ denotes the Sun. While in the central parts 
of this galaxy there exists good agreement with Newtonian model, in the 
outskirts, observed velocities do not show Keplerian fall-off. It is assumed 
that the mass in these outer parts is dominated by low-luminosity material -- 
a dark halo.

This material could be in two forms: baryonic and nonbaryonic. According to 
Persic and Salucci (1996), (hereafter PS96):

$$\Omega ^{\rm gal} _{\rm bar}=\Omega ^{\rm E} _{\rm bar}+\Omega ^{\rm S} 
_{\rm bar}=2\times 10^{-3},$$

\noindent where $\Omega _{\rm bar}={\rho _{\rm bar}\over \rho _{\rm crit}}$. 

PS96 found that spiral (S) and elliptical  (E) galaxies contribute the same 
cosmological stellar mass density. However, recent measurements imply that the 
density parameter $\Omega _0=1$  (with estimated 
standard error $\pm 0.2$) thus indicating the existence of non-baryonic DM 
(Kolb and Turner, 1993).

If we want to make the model of a spiral galaxy (for example, NGC3198) we 
could use the following Richstone and Tremaine (1986) approach for the mass 
density:

$$\rho (r) = {\rho _c\over (1+{r^2\over r_c^2})^{3\over 2}}.$$

\noindent where $r _c$ is the core radius. For small values of $r$, i.e. 
$r\ll r_c$:

$$\rho=\rho _c (1-{3\over 2} {r^2\over r_c^2}+ \cdots ).$$

The velocity distribution is isotropic and independent of position, the 
mass-to-light ratio is independent of radius. In this model there exists a 
well defined, flat central core. The velocity dispersion is $\sigma =
<v^2>^{1\over 2}$ in one dimension, the pressure is $p=\rho \sigma ^2$ and the 
pressure per unit volume is $-{\partial p\over \partial r}$. The condition
for gravitational 
equilibrium is: 
$${\partial  p\over \partial  r}={3\rho_c\sigma ^2r\over r_c^2}={GM(r)\over 
r^2}\rho= {4\pi\over 
3}G\rho_c^2r.$$

Thus, one gets the central mass density:

$$\rho _c={9\sigma ^2\over 4\pi G r_c^2}$$

(Peebles, 1993, Padmanabhan, 1993). 

Schramm and Steigman (1981) considered relic neutrinos with mass $m_\nu {}
\gsim {1\over 2}$ eV and obtained the same result.
According to this paper neutrinos could have collapsed gravitationally during 
the formation of astrophysical systems whose potential wells are sufficiently 
deep:

$${GM(\le r_c)\over r_c}\approx 3\sigma ^2 \gsim v^2_{\nu}.$$

Following Peebles' (1993) reasoning, one can establish a possible model for 
the present mean distribution of neutrinos in a dark halo via spherically 
symmetric isothermal form:

$${\cal N}_f={\cal N}_0 e^{-{1\over \sigma^2} ({v^2\over 2}+ \varphi 
(r))}\eqno (2)$$

\noindent where ${\cal N}_f$ is defined as:

$${\cal N}_f=\bigl \langle {1\over e^{pc\over kT_\nu} +1}\bigr \rangle , \, \, 
\, T_\nu = ({4\over 11})^{1\over 3}T_0 (1+z)$$ 

\noindent where $T_0$ is the present cosmic microwave background temperature, 
$T_0=2.73\pm 0.01$ K (Smooth and Other, 1995). In the equation (2) the 
gravitational potential per unit mass at radius $r$ is $\varphi (r)$, with 
$\varphi (0)=0$, ${\cal N}_0 \le 0.5$.

This distribution  can be used in order to find the mean density:

$$\rho _\nu (r) ={2m_\nu \over (2\pi \hbar )^2} \, \int {\cal N}_f d^3 p=
{{\cal N}_0m_\nu ^4 \sigma ^3 \over 2^{1\over 2}\pi ^{3\over 2} \hbar ^3}
e^{-{\varphi (r)\over \sigma ^2}} \eqno(2a)$$

\noindent (Peebles, 1993).

After solving the Poisson equation for the gravitational potential:

$$\nabla ^2 \varphi = {d^2 \varphi\over dr^2}+ {2\over r}{d \varphi\over 
dr}=4\pi G \rho _\nu (r)$$

\noindent one obtains, for $r\gg \alpha$, $\alpha ^2={\pi ^{1\over 2}\over 
2^{3\over 2}} {\hbar ^3\over G{\cal N}_0 m_\nu ^4 \sigma}$:

$$\rho_\nu (r)={\sigma ^2 \over 2\pi G r^2}=
{v ^2 \over 6\pi G r^2}.\eqno(3)$$

We put in  the value for $v^2$, i.e. the mean square 
velocity $<v^2>$, $<v^2>=3\sigma ^2$ (Padmanabhan, 1993).

It follows from equations (2a) and (3) that the mass is equal to: 

$$m_\nu ^4={1\over 6\pi} {3\overwithdelims () 2\pi}^{1\over 2} {h^3\over 
Gvr^2_c},\eqno(4)$$

\noindent where we have assumed that ${\varphi (r)\over \sigma ^2}\rightarrow
0$ and ${\cal N}_0=0.5$.

If one inserts the values characteristic for the Milky Way ($v\sim 230\, {\rm 
km\over s}$ and $r_c \sim 8$ kpc), one obtains the following result for the 
neutrino mass:

$$m_\nu \approx 27\, {\rm eV}.$$

This value is the {\it lower\/} limit for $m_\nu$, while the {\it upper\/} 
limit is given in the equation (1).
The obtained value plays the crucial role in the decaying dark matter (DDM) 
theory, firstly proposed by Melott (1984) and later developed by Sciama 
(1990a, 1990b,  1993). According to this theory the mass of the tau neutrino is:

$${m_\nu}_\tau=29.21\pm 0.15\, {\rm eV}.$$

Melott {\it et al\/}. (1994) while considering decaying neutrinos in galaxy 
clusters obtained the lower limit for the neutrino lifetime $\tau  _{23}$
in the units $10^{23}$ s:
$\tau _{23} > (3\pm 1) {29\, {\rm eV}\overwithdelims () m_\nu}.$
Such a decay could be observed
(e.g., Samurovi\'c and \v Celebonovi\'c, 1995). Experiment that has 
been proposed, EURD, will have to prove the 
existence of a decay line derived from the photons with energy $\sim 15$ 
eV.\footnote {$^1$}  {\tt URL: http://www.laeff.esa.es/eng/laeff/activity/eurd.html}
\bigskip
\leftline {\bf 3. Rotational Curves of Spiral Galaxies}

\medskip

PSS presented the profiles of 134 RCs (references on the measurements can be 
found therein). We used the equation (4) which after 
inserting the appropriate values becomes:

$$m_\nu \, {\rm [eV]}={299.0362 \over v_{\rm [km/ s]}^{1\over 4}\, r_{c{\rm 
[kpc]}}^{1\over 2}}. \eqno(5)$$

\noindent For the estimation of the core radius, $r_c$, we used Kormendy's 
relation (Ashman, 1992) according to which:
$$r_c \approx 5.9 {L_B\overwithdelims () 10^9 L_{B\odot}}^{0.34}\, {\rm [kpc]}.
\eqno (6)$$
In the Table (1) we present the values from PSS together with estimated 
values for the neutrino mass (most probably of the tau neutrino). Although the 
most important galactic parameters are not known accurately (Spergel, 1996), 
one can see  that the values of the neutrino mass 
are gathered within  the range $20\lsim m_\nu \lsim 30\, {\rm eV}$;
the mean value is $22\pm 7$ eV.
The obtained result could lead to the conclusion that massive neutrinos {\it 
could\/} dominate in the spiral galaxies, but additional measurements of 
galaxies' parameters and laboratory measurements of neutrino mass are 
indispensible. 

\bigskip
\noindent {\bf Acknowledgments}
\medskip
S.S. would like to thank A.L. Melott for the explanation of the path 
of formation of the DDM theory and R.J. Splinter for the description of the 
serious problems of this theory.

\vfill\eject

\bigskip

\bigskip

{\bf REFERENCES}
\bigskip

\item{}\kern-\parindent{Ashman, K. M.: 1992, {\it Pub A.S.P.\/}, 
{\bf 104}, 1109.}

\item{}\kern-\parindent{Athanassopoulos, C. {\it et al\/}.: 1995, {\it Phys. 
Rev. Lett.}, {\bf 75}, 2650.}

\item{}\kern-\parindent{Hill J.E.: 1995, {\it Phys. 
Rev. Lett.}, {\bf 75}, 2654.}

\item{}\kern-\parindent{Kolb, E.W. and Turner, M.S.: 1993, {\it The Early 
Universe} (paperback edition), Addison Wesley.}

\item{}\kern-\parindent{Melott A.L.: 1984, {\it Astronomicheskii
Zhurnal}, {\bf 61}, 817; {\it Soviet Astronomy},  {\bf 28},
478}. 

\item{}\kern-\parindent{Melott A.L., Splinter R.J., Persic, M. and Salucci, 
P.: 1994, {\it Ap.J.}, {\bf 421}, 16}.

\item{}\kern-\parindent{Montanet L. {\it et al.\/}, Physical Review {\bf D50}, 
1173 (1994)   and 1995 off-year partial update for the 1996 edition available
  on  the PDG WWW pages (URL: {\tt http://pdg.lbl.gov/}).}

\item{}\kern-\parindent{Padmanabhan, T.: 1993, {\it Structure Formation in the 
Universe}, Cambridge University  Press.}

\item{}\kern-\parindent{Peebles, P.J.E.: 1993, {\it Principles of Physical 
Cosmology}, Princeton University Press.}

\item{}\kern-\parindent{Persic, M., Salucci, P. and Stel, F. (PSS): 1995, 
astro-ph/9506004 preprint.}

\item{}\kern-\parindent{Persic, M., Salucci, P.: (PS96): 1996, 
astro-ph/9601018 preprint.}

\item{}\kern-\parindent{Primack, J.R.: 1996,  astro-ph/9604184 preprint.}

\item{}\kern-\parindent{Richstone, D. O. and Tremaine, S. 1986, {\it A.J.}, 
{\bf 92}, 72.}

\item{}\kern-\parindent{Rubin, V. C., Burstein, D., Ford, W. K.  and Thonnard 
N. 1985, {\it Ap.J.}, {\bf 289}, 81.}

\item{}\kern-\parindent{Samurovi\'c, S. and \v Celebonovi\'c, V.: 1995, {\it 
Publ. Obs. Astron. Belgrade}, {\bf 50}, 121. (astro-ph/9512029)}

\item{}\kern-\parindent{Sarkar, S.: 1996, hep-ph/9602260 preprint.}

\item{}\kern-\parindent{Sciama, D. W.: 1990a, {\it Ap.J.}, {\bf 364}, 549.}

\item{}\kern-\parindent{Sciama, D. W.: 1990b, {\it M.N.R.A.S.}, {\bf 244}, 
9p.}
         
\item{}\kern-\parindent{Sciama, D. W.: 1993, {\it Modern Cosmology and the 
Dark Matter Problem}, Cambridge University  Press.}

\item{}\kern-\parindent{Schramm, D. N. and Steigman G. 1981, {\it Ap.J.}, 
{\bf 243}, 1.}

\item{}\kern-\parindent{Smooth, G.F. and Other, A.N.: 1995, astro-ph/9505139
preprint.}

\item{}\kern-\parindent{Spergel, D.: 1996, astro-ph/9603026 preprint.}

\item{}\kern-\parindent{van Albada, T. S., Bahcall, J. N., Begeman, K.
and Sancisi, R. 1985, {\it Ap.J.}, {\bf 295}, 305.}

\vfill\eject

\vbox{\tabskip=0pt \offinterlineskip
\def\podvuci{\noalign{\hrule}}
\def\razmak{\noalign{\vskip.1cm}}
\halign to 12cm{\strut#& \vrule#\tabskip=0pt  plus 10pt minus5pt &
\hfil#\hfil&\vrule#& 
 \hfil#&\vrule#& \hfil#&\vrule#& \hfil#&\vrule#& \hfil#&\vrule# width2pt& 
\hfil#\hfil&\vrule#&
 \hfil#&\vrule#&  \hfil#&\vrule#& \hfil#&\vrule#&    \hfil#&\vrule#
\tabskip=0pt\cr\podvuci
&& {\rm Name}&& $M_B$&&\ $v$ [${\rm km\over s}$] && $\ r_c\ {\rm [kpc]}$
&& $m_\nu$ {\rm [eV]} 
&& {\rm Name}&& $M_B$&&\ $v$ [${\rm km\over s}$]&& $\ r_c\ {\rm [kpc]}$  
&& $m_\nu$ {\rm [eV]} 
&\cr\podvuci\razmak\podvuci

&&N55&&\ -17.83 &&80 &&7.6 &&36.26 &&U 3269&&\ -21.01&&185 &&20.58 &&17.87 &\cr 

&&N224&&\ -20.81 &&250 && 19.33 &&17.10 &&U 3282&&\ -21.57&&220 &&24.53 &&15.68 
&\cr 

&&N247&&\ -17.52 &&110 && 6.9 &&28.62 &&U 4375&&\ -19.59&&190 &&13.19 &&22.17 
&\cr 

&&N253&&\ -19.82 &&215 &&14.17 &&20.74 &&U 11810&&\ -20.39&&185 &&16.95 
&&16.96 &\cr 

&&N300&&\ -16.83 &&90 &&4.29 &&46.89 &&U 12417&&\ -19.45&&126 &&12.63 &&25.12 
&\cr 

&&N598 &&\ -18.31 &&108 &&8.84 &&31.20 &&U 12533&&\ -19.93&&210 &&14.68 
&&20.51 &\cr 

&&N628&&\ -20.12 &&200 &&15.58 &&20.15 &&U 12810&&\ -21.45&&220 &&23.62 
&&15.98 &\cr 

&&N697&&\ -21.16 &&214 &&15.77 &&19.69 &&27-008&&\ -20.48&&171 &&17.44 &&19.80 
&\cr 

&&N753 &&\ -21.54 &&215 &&24.30 &&15.84 &&30-009&&\ -21.71&&300 &&25.63 
&&14.19 &\cr 
                                                             
&&N801&&\ -21.26 &&220 &&22.26 &&16.46 &&40-012&& ... &&200 && ... && ...  
&\cr 

&&N891 &&\ -20.50 &&230 &&17.54 &&18.33 &&41-009&& ... &&198 && ... && ...  
&\cr 

&&N925 &&\ -19.24 &&116 &&11.82 &&26.50 &&69-011&& ... &&177 && ... &&  ...  
&\cr 

&&N1035&&\ -19.04 &&125 &&11.10 &&26.84 &&71-005&&\ -20.81&&240 &&19.33 && 
17.28 &\cr 

&&N1085 &&\ -21.92 &&302 &&27.37 &&13.71 &&75-037&&\ -19.06&&111 &&11.18 
&&27.55 &\cr 

&&N1090&&\ -20.72 &&173 &&18.80 &&19.02 &&82-008&&\ -21.04&&255 &&20.78 
&&16.42 &\cr 

&&N1097 &&\ -20.71 &&250 &&18.74 &&17.37 &&88-016&&\ -20.89&&215 &&19.82 
&&16.81 &\cr 

&&N1114   && ... &&198 &&... &&... &&116-012&&\ -17.75&&134 &&7.42 &&32.28 
&\cr 

&&N1247 &&\ -21.07 &&270 &&20.97 &&16.11 &&121-006&&\ -18.46&&129 &&9.26 
&&29.16 &\cr 

&&N1365&&\ -20.27 &&275 &&16.32 &&18.17 &&123-023&&\ -19.62&&150 &&13.32 
&&23.41 &\cr 

&&N1417 &&\ -21.15 &&245 &&15.72 &&19.06 &&141-020&&\ -21.01&&239 &&20.58 
&&16.76 &\cr 

&&N1560 &&\ -16.80 &&62 &&5.5 &&45.41 && 141-034&&\ -21.40&&277 &&23.25 
&&15.20 &\cr 

&&N1832 &&\ -20.25 &&180 &&16.22 &&20.27 &&184-051&&\ -21.40&&240 &&23.26 
&&15.75 &\cr 

&&N2336&&\ -21.79 &&250 &&26.27 &&14.67 &&215-039&&\ -20.82&&160 &&19.39 
&&19.09 &\cr 

&&N2403 &&\ -19.24 &&127 &&11.82 &&25.90 &&235-016&&\ -20.74&&202 &&18.91 
&&18.24 &\cr 

&&N2558 &&\ -20.43 &&245 &&17.16 &&18.24 &&240-011&&\ -20.53&&222 &&17.71 
&&18.41 &\cr 

&&N2595&&\ -21.15 &&300 &&21.50 &&15.49 &&269-019&&\ -20.85&&192 &&19.58 
&&18.16 &\cr 

&&N2742 &&\ -20.10 &&170 &&15.48 &&21.05 &&282-003&&\ -21.49&&198 &&23.92 
&&16.30 &\cr 

&&N2841 &&\ -21.83 &&320 &&26.61 &&13.71 &&284-024&&\ -20.45&&165 &&17.27 
&&20.08 &\cr 

&&N2903 &&\ -20.52 &&206 &&17.65 &&18.79 &&286-016&&\ -20.20&&171 &&15.97 
&&20.69 &\cr 

&&N2998 &&\ -21.16 &&215 &&21.57 &&16.81 &&287-013&&\ -20.55&&167 &&17.82 
&&19.71 &\cr 

&&N3109 &&\ -16.43 &&55 &&4.90 &&49.58 &&289-010&&\ -18.88&&98 && 10.56 
&&29.24 &\cr 

&&N3145&&\ -21.22 &&275 &&21.98 &&15.66 &&299-004&&\ -20.96&&187 &&20.26 
&&17.96 &\cr 

&&N3198 &&\ -20.17 &&156 &&15.82 &&21.27 &&306-032&&\ -20.24&&179 &&16.17 
&&20.33 &\cr 

&&N3200 &&\ -21.09 &&280 &&21.10 &&15.91 &&322-045&&\ -19.35&&165 &&12.23 
&&23.85 &\cr 

&&N3223 &&\ -21.76 &&253 &&26.03 &&14.70 &&322-076&&\ -20.24&&175 &&16.17 
&&20.44 &\cr 

&&N3992 &&\ -21.67 &&275 &&25.31 &&14.60 &&346--014&&\ -18.60&&98 &&9.68 
&&30.55 &\cr 

&&N4013 &&\ -19.73 &&196 &&13.79 &&21.53 &&347-033&&\ -19.84&&198 &&14.27 
&&21.10 &\cr 

&&N4062 &&\ -19.47 &&160 &&12.71 &&23.59 &&350-023&&\ -21.08&&230 &&21.04 
&&16.74 &\cr 

&&N4236&&\ -18.74 &&88 &&10.11 &&30.71 &&352-053&&\ -20.44&&250 &&17.22 
&&18.12 &\cr 

&&N4258 &&\ -20.73 &&205 &&18.85 &&18.20 &&374-027&&\ -21.35&&258 &&22.89 
&&15.60 &\cr 

&&N4348 &&\ -19.52 &&188 &&12.91 &&22.48 &&376-002&&\ -20.43&&200 &&17.16 
&&19.19 &\cr 

&&N4565&&\ -20.86 &&240 &&19.64 &&17.14 &&379-006&&\ -19.87&&169 &&14.40 
&&21.85 &\cr 

&&N4605&&\ -16.84 &&100 &&5.58 &&40.04 &&383-002&&\ -20.54&&190 &&17.77 
&&19.11 &\cr 
 \podvuci}} 

\vfill\eject

\vbox{\tabskip=0pt \offinterlineskip
\def\podvuci{\noalign{\hrule}}
\def\razmak{\noalign{\vskip.1cm}}
\halign to 12cm{\strut#& \vrule#\tabskip=0pt  plus 10pt minus5pt &
\hfil#\hfil&\vrule#& 
 \hfil#&\vrule#& \hfil#&\vrule#& \hfil#&\vrule#& \hfil#&\vrule# width2pt& 
\hfil#\hfil&\vrule#&
 \hfil#&\vrule#&  \hfil#&\vrule#& \hfil#&\vrule#&    \hfil#&\vrule#
\tabskip=0pt\cr\podvuci
&& {\rm Name}&& $M_B$&&\ $v$ [${\rm km\over s}$] && $\ r_c\ {\rm [kpc]}$
&& $m_\nu$ {\rm [eV]} 
&& {\rm Name}&& $M_B$&&\ $v$ [${\rm km\over s}$]&& $\ r_c\ {\rm [kpc]}$  
&& $m_\nu$ {\rm [eV]} 
&\cr\podvuci\razmak\podvuci

&&N4682&&\ -19.58 &&175 &&13.15 &&22.67 &&383-088&&\ -20.04&&180 &&15.19 
&&20.95 &\cr 

&&N4800&&\ -18.76 &&172 &&10.17 &&25.89 &&437-030&&\ -20.46&&205 &&17.33 
&&18.99 &\cr 

&&N5033 &&\ -20.67 &&218 &&18.50 &&18.09 &&439-018&&\ -21.71&&245 &&25.63 
&&14.93 &\cr 

&&N5055 &&\ -20.65 &&215 &&18.39 &&18.21 &&439-020&&\ -19.91&&215 &&14.58 
&&20.45 &\cr

&&N5371 &&\ -21.72 &&240 &&25.71 &&14.98 &&444-047&&\ -19.44&&145 &&12.59 
&&24.29 &\cr 

&&N5585 &&\ -18.47 &&80 &&9.29 &&32.80 &&444-086&&\ -19.95&&210 &&14.77 
&&20.44 &\cr 

&&N5673 &&\ -20.30 &&138 &&16.48 &&21.49 &&445-058&&\ -20.72&&205 &&18.80 
&&18.23 &\cr 

&&N5905 &&\ -20.87 &&235 &&19.70 &&17.21 &&446-044&&\ -19.22&&148 &&11.75 
&&25.01 &\cr 

&&N5907 &&\ -20.69 &&225 &&18.62 &&17.89 &&481-002&&\ -18.22&&119 &&8.59 
&&30.89 &\cr

&&N6503&&\ -18.62&&122 &&9.74 &&28.83 && 499-005&&\ -20.84&&170 &&19.52 
&&18.75 &\cr 

&&N6674&&\ -21.33&&255 &&22.75 &&15.69 &&502-002&&\ -20.25&&210 &&16.22 
&&19.50 &\cr 

&&N7083&&\ -21.27&&220 &&22.33 &&16.43 && 507-007&&\ -21.10&&263 &&21.17 
&&16.14 &\cr

&&N7331&&\ -20.92&&243 &&20.01 &&16.93 && 509-091&&\ -20.03&&150 &&15.14 
&&21.96 &\cr 

&&N7339&&\ -19.21&&170 &&11.73 &&24.20 && 533-004&&\ -19.23&&150 &&11.79 
&&24.89 &\cr 

&&N7536&&\ -20.18&&183 &&15.87 &&20.41 && 543-012&& ... &&174 && ... && ...  
&\cr 

&&N7591&&\ -20.34&&200 &&16.69 &&19.47 &&548-032&&\ -17.58&&66 &&7.03 &&39.57 
&\cr 

&&N7593&&\ -19.61&&150 &&13.28 &&23.45 &&555-016&&\ -20.19&&222 &&15.92 
&&19.42 &\cr 

&&N7606&&\ -21.15&&265 &&21.50 &&15.98 && 563-014&& ... &&148 && ... && ...  
&\cr 

&&N7631&&\ -20.18&&198 &&15.87 &&20.01 && 564-020&&\ -19.00&&92 &&10.97 
&&29.16 &\cr 

&&N7793&&\ -17.79&&111 &&7.51 &&33.62 && 566-022&&\ -18.91&&138 &&10.66 
&&26.72 &\cr 

&&I 467&&\ -19.77&&150 &&13.96 &&22.87 && 601-009&&\ -21.55&&258 &&24.37 
&&15.11 &\cr 

&&I 2974&& ... &&230 && ... && ... && M-3-1042&& ... &&148 && ... && ...  &\cr 

&&U 2259&&\ -16.54&&81 &&5.08 &&44.24 && && && && &&  &\cr
\podvuci}} 
    
\bigskip
\noindent{Table 1. Values of radius $r$ in kpc, velocity $v$ in ${\rm km / s}$ 
and mass of neutrino $m_\nu$ for 134 spiral galaxies.}

\bye